# An ultra-low-power CGRA for accelerating Transformers at the edge

Rohit Prasad
Université Paris-Saclay, CEA, List, F-91120, Palaiseau, France, [firstname].[lastname]@cea.fr

*Abstract*—Transformers have revolutionized deep learning with applications in natural language processing, computer vision, and beyond. However, their computational demands make it challenging to deploy them on low-power edge devices. This paper introduces an ultra-low-power, Coarse-Grained Reconfigurable Array (CGRA) architecture specifically designed to accelerate General Matrix Multiplication (GEMM) operations in transformer models tailored for the energy and resource constraints of edge applications. The proposed architecture integrates a $4 \times 4$ array of Processing Elements (PEs) for efficient parallel computation and dedicated $4 \times 2$ Memory Operation Blocks (MOBs) for optimized LOAD/STORE operations, reducing memory bandwidth demands and enhancing data reuse. A switchless mesh torus interconnect network further minimizes power and latency by enabling direct communication between PEs and MOBs, eliminating the need for centralized switching. Through its heterogeneous array design and efficient dataflow, this CGRA architecture addresses the unique computational needs of transformers, offering a scalable pathway to deploy sophisticated machine learning models on edge devices.

## I. Introduction

The rapid advancements in deep learning (DL) have ushered in a new era of intelligent applications across domains such as natural language processing, computer vision, and edge computing [14]. Transformers, in particular, have emerged as a breakthrough architecture for tasks that require high accuracy in complex pattern recognition, including language translation, image classification, and autonomous decision-making [17]. Despite their transformative capabilities, the computational demands of these models pose a significant challenge, particularly for real-time and power-constrained environments such as edge devices. This demand for efficient processing has sparked substantial interest in energy-efficient hardware accelerators.

CGRA architectures offer a promising solution to address these computational challenges [1], [3], [12], [16]. CGRAs are characterized by arrays of programmable PEs and an interconnected network, enabling parallelism and data movement optimizations [11]. Such architectures can suit the unique requirements of ML workloads. Unlike general-purpose processors (GPPs) or even graphics processing units (GPUs), CGRAs provide a flexible yet efficient approach to accelerate the dense and irregular matrix operations that underpin transformer models, particularly the GEMM operations integral to multi-layer perceptron and attention mechanisms in transformers. By supporting programmability, CGRAs allow for the adaptability required in diverse DL applications while offering greater energy efficiency than traditional architectures.

Recent advancements in CGRA design have introduced innovations such as heterogeneous arrays of PEs and optimized interconnect topologies tailored to the dataflow patterns of deep neural network (DNN) models [4]. Such improvements hold significant potential for accelerating DL workloads while meeting the stringent power and area constraints of edge computing devices. Moreover, as transformer models grow in complexity, CGRAs offer a scalable architecture capable of handling the increasing demand for parallel computation and memory bandwidth.

This paper explores the application of CGRAs for accelerating transformer workloads, with a focus on GEMM operations. The author presents an analysis of how CGRA architectures can be tailored to these workloads, highlighting recent innovations in PE design, memory management, and interconnect strategies. The study demonstrates that with these optimizations, CGRAs can achieve substantial improvements in power efficiency and performance, making them a competitive solution for DL acceleration in edge environments.

The rest of the paper is organized as follows: section II discusses related work and background, section III discusses the proposed CGRA architecture, section IV gives an overview of the GEMM acceleration for transformers, and finally section V provides a conclusion.

## II. Related Work and Background

The exponential growth of machine learning (ML) applications, particularly those involving DNNs and transformer models, has intensified the demand for efficient hardware accelerators capable of supporting their computational and memory-intensive workloads. For edge computing applications, this need is compounded by the constraints of limited power and area resources, which often make GPPs and GPUs impractical. CGRA architectures have emerged as a promising alternative due to their ability to deliver high performance and energy efficiency alongside programmable flexibility. CGRAs combine hardware-like efficiency with the adaptability of software-based solutions, making them especially suitable for diverse ML workloads and signal-processing tasks.

## A. CGRA Suitability for ML Workloads

CGRAs are particularly well-aligned with the needs of ML and DNN workloads for several reasons. First, their architectural design supports high-throughput computation, enabling efficient handling of large datasets and complex arithmetic operations typical in DNNs [9]. CGRAs utilize arrays of programmable PEs connected by configurable interconnects, allowing parallelism to be tailored to specific workloads, such as the extensive matrix multiplications in transformer models [2]. Additionally, CGRAs' dataflow-oriented approach is beneficial for ML applications, as data is processed based on availability rather than strict control flow [7]. This closely matches the structure of many DNN and transformer algorithms.

## B. Applications of CGRAs in Signal Processing and Deep Learning

The potential of CGRAs has been demonstrated in applications spanning both signal processing and deep learning [12]. For instance, CGRAs excel at stream processing, a capability that is advantageous for image and video processing tasks. In industry, Samsung utilizes CGRA-based platforms for high-resolution video processing in its 8K UHD displays, highlighting CGRAs' ability to meet stringent performance requirements for real-time applications [5]. Beyond traditional signal processing, CGRAs have also proven effective in accelerating DNN workloads. Architectural extensions such as NP-CGRA leverage CGRAs' flexibility to minimize data movement, which in turn reduces energy consumption—a critical factor for ML applications at the edge [6].

To support efficient mapping of ML operations, CGRA designs often incorporate features like systolic computation, which allows for parallel, structured processing of DNN layers, such as convolutional layers in convolutional neural networks (CNNs). These features enable CGRAs to achieve data-level parallelism and exploit reusable data patterns, significantly improving the execution efficiency of DNN operations compared to single-instruction multiple-data (SIMD) and single-instruction multiple-thread (SIMT) architectures [15].

## C. CGRAs for Edge and Low-Power IoT Devices

For edge devices and low-power IoT applications, CGRAs offer an attractive balance between flexibility and energy efficiency. Unlike fixed-function accelerators, CGRAs provide tunable precision and reconfigurability, supporting both integer and floating-point operations to accommodate a range of ML and signal processing tasks. Recent architectures, such as TRANSPIRE, incorporate transprecision capabilities and multiple-instruction multiple-data (MIMD) structures to deliver scalable performance for ultra-low-power applications, enabling edge devices to handle ML computations that would typically require offloading to more powerful, energy-intensive systems [12].

## D. Architectural Innovations in CGRA Design

Ongoing research in CGRA design is producing architectures that address several core challenges in ML and DNN acceleration, including efficient memory access, programmability, and multi-level parallelism. Trends in CGRA architecture include:

*1) Programming-Driven Design:* To make CGRAs more accessible to developers, recent designs prioritize programmability through advanced software tools, parallel programming models, and domain-specific languages (DSLs) that expose CGRA features while abstracting hardware complexity. This shift facilitates the deployment of ML algorithms on CGRAs without deep hardware expertise [13].

*2) Multilevel Parallel Computation:* CGRA designs increasingly incorporate multi-level parallelism, spanning instruction-level parallelism, data-level parallelism, and task-level parallelism [15]. Techniques such as dataflow and distributed control schemes enable concurrent execution across PEs, optimizing resource utilization and supporting complex ML workflows.

*3) Efficient Memory Access and Near-Memory Computing:* Memory bandwidth is a significant bottleneck for ML tasks, especially in data-intensive layers of DNNs. CGRAs address this by incorporating vectorized or streaming memory interfaces and programmable memory-managing units to improve data access efficiency. Advanced CGRA designs also explore 3D chip stacking, placing PEs closer to memory through near-memory computing strategies, which can increase bandwidth while reducing data transfer distances. For example, high-bandwidth memory (HBM) integration with CGRAs allows rapid data access and efficient processing for real-time ML applications [8], [10].

These advances reflect a growing trend toward making CGRAs versatile yet specialized enough to meet the specific demands of ML workloads. As transformer and DNN models continue to grow in complexity, CGRA-based accelerators that combine programmability with efficient parallelism and memory management stand out as a promising solution for edge-based AI applications.

## III. CGRA ARCHITECTURE

### A. CGRA Integrated System

Figure 1 illustrates the integrated system architecture, where the CGRA subsystem is loosely coupled with a host CPU subsystem. Data exchange between the CPU and CGRA subsystems occurs through a shared L1 memory accessible via an interconnect. The CGRA subsystem includes a 4KiB Context Memory, a Memory Controller, and the CGRA. The Memory Controller retrieves and interprets configuration data from the Context Memory, distributing instructions across each PE and MOB in the CGRA array. This setup ensures that all components are pre-configured before initiating kernel execution on the CGRA.

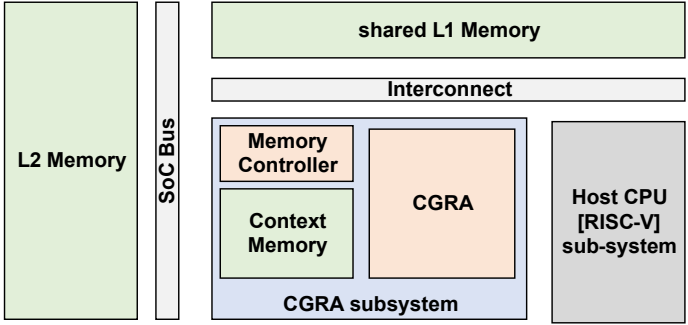

Fig. 1. CGRA integrated system

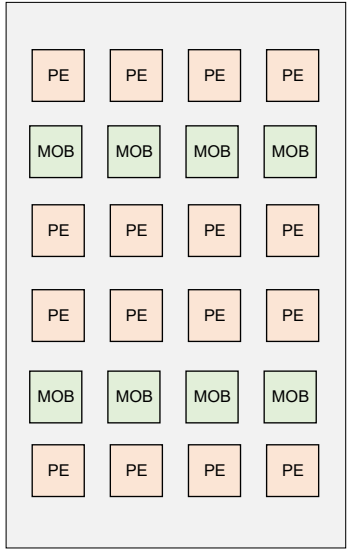

Fig. 2. PE and MOB array

### B. Heterogeneous Array Design

*1) Processing Element Array:* A core innovation of this architecture is the deployment of a 4×4 PE array (Figure 2) tailored for low-latency arithmetic operations required by GEMM, a key component in transformer workloads. Each PE in this array is designed to perform fundamental arithmetic tasks (i.e., dot-product by incorporating additions and multiplications on packed data) in parallel, which enables efficient sub-matrix multiplication. This approach allows the GEMM workload to be split into smaller, manageable units that are distributed across the PEs in the 4 × 4 grid, thereby achieving fine-grained parallelism. This capability is particularly beneficial for transformers, as it accelerates both the self-attention and feedforward layers that rely heavily on matrix operations.

Key Benefits:

- *Parallel Computation for GEMM*: By distributing sub-matrix operations across the PEs, the architecture supports concurrent execution of GEMM blocks, significantly speeding up transformer inference.
- *Optimized Data Processing*: The dedicated design of each PE for arithmetic operations ensures high computational efficiency, which is essential for sustaining the heavy workloads of transformers without overburdening memory or communication resources.

*2) Memory Operation Block Array:* In addition to the PEs, the architecture introduces a 4 × 2 MOB array (Figure 2) explicitly designed for LOAD/STORE operations. This MOB handles data transfer between external memory and the PE array, ensuring that each PE has prompt access to the required data. By dedicating a separate MOB for memory operations, the architecture effectively separates data retrieval/storage tasks from computational tasks, minimizing contention and improving overall data access efficiency.

Key Benefits:

- *Enhanced Memory Bandwidth Utilization*: By managing memory tasks in a dedicated block, the architecture reduces bottlenecks associated with data access, a critical need for transformer models that require frequent memory retrievals.
- *Reduced PE Idle Time*: The separation of LOAD/STORE and computational operations minimizes data stalling, as data can be prefetched and made available to the PEs without disrupting ongoing computations. This balance between computation and memory management enables sustained high throughput for GEMM execution.

### C. Switchless Mesh Torus Interconnect

A unique feature of this CGRA is its switchless mesh torus interconnect, which facilitates data transfers between the PEs and MOBs without a conventional switching network. Traditional interconnects rely on centralized switches, which introduce latency and add to power consumption as data packets are routed across the network. By adopting a switchless mesh torus interconnect, this architecture minimizes these drawbacks, providing direct communication between neighboring PEs. The torus topology, which wraps around the grid edges, allows data to take shorter paths, conserving energy and reducing latency.

Key Benefits:

- *Reduced Dynamic Power Consumption*: Removing centralized switches reduces the dynamic power cost typically associated with packet routing and switching, a significant advantage for energy-constrained edge devices.
- *Predictable Data Flow*: The switchless interconnect establishes fixed, predictable paths for data transfer, which complements the structured dataflow of GEMM operations in transformers. This design choice enhances the efficiency of GEMM, as data routing aligns well with the patterns required by transformer models.

## IV. GEMM ACCELERATION FOR TRANSFORMERS

### A. GEMM Computation Strategy

*1) Block-Wise GEMM Execution:* General Matrix Multiplication (GEMM) is a central operation in many ML

$$\begin{bmatrix} a_{11} & a_{12} & \cdots & a_{1n} \\ a_{21} & a_{22} & \cdots & a_{2n} \\ \vdots & \vdots & \ddots & \vdots \\ a_{m1} & a_{m2} & \cdots & a_{mn} \end{bmatrix} \times \begin{bmatrix} b_{11} & b_{12} & \cdots & b_{1p} \\ b_{21} & b_{22} & \cdots & b_{2p} \\ \vdots & \vdots & \ddots & \vdots \\ b_{n1} & b_{n2} & \cdots & b_{np} \end{bmatrix} = \begin{bmatrix} c_{11} & c_{12} & \cdots & c_{1p} \\ c_{21} & c_{22} & \cdots & c_{2p} \\ \vdots & \vdots & \ddots & \vdots \\ c_{m1} & c_{m2} & \cdots & c_{mp} \end{bmatrix}$$

$$c_{11} = a_{11}b_{11} + a_{12}b_{21} + \cdots + a_{1n}b_{n1} = \sum_{k=1}^{n} a_{1k}b_{k1}$$

Fig. 3. Matrix multiplication of $A$ and $B$ resulting in matrix $C$. The row in matrix $A$, column in matrix $B$, and element in matrix $C$ are highlighted with red-bordered boxes to illustrate the computation of $c_{11}$ as the dot product of the row and column.

models, including transformer architectures, due to its use in processing the large matrix operations involved in attention and feedforward neural network layers. In this architecture, GEMM operations are executed in a block-wise manner, breaking down the matrix multiplications into smaller, manageable sub-matrices. This strategy aligns well with the $4 \times 4$ PE array in the CGRA, enabling parallel computation of these blocks across the array.

Dividing matrices into sub-blocks for GEMM execution provides multiple advantages:

- *Increased Data Reuse*: By keeping data within the PE array for as long as possible, this approach reduces the need for frequent external memory accesses, which are typically costly in terms of latency and energy. The sub-matrix blocks allow each PE in the $4 \times 4$ array to repeatedly use data from the previous cycles, capitalizing on data locality.
- *Reduced Memory Bandwidth Requirements*: Since the PEs can operate on data blocks held in the PE array, there is a reduced reliance on external memory, thereby decreasing memory bandwidth demands. This efficiency is particularly valuable for edge devices, where memory bandwidth is often limited and must be managed to minimize power consumption.

*2) Memory Management:* The efficient handling of data LOAD/STORE operations is critical to maximizing the throughput of GEMM operations, especially in resource-constrained environments. The CGRA uses a $4 \times 2$ MOB array dedicated to managing these operations. The MOBs efficiently serve as a bridge between the PEs and external memory, allowing rapid access to data needed for transformer parameters and intermediate computations.

Key aspects of this memory management strategy include:

- *Rapid Data Access*: Each PE in the $4 \times 4$ array can quickly access the required data through the MOB array, reducing the wait time associated with data retrieval and enabling smoother, more continuous data flow.
- *Reduced Data Stalling*: By alternating LOAD/STORE operations in the MOB with computation in the PEs, the architecture minimizes instances of data stalling. Data stalling, a common issue in matrix-intensive workloads, occurs when computation is delayed due to data not being readily available. The efficient interleaving of memory and ALU operations mitigates this problem, maximizing GEMM processing efficiency and reducing idle time for PEs.

Together, these elements allow block-wise GEMM computation to maintain high throughput and low latency, essential for real-time transformer processing in edge scenarios.

*B. Optimizing Transformer Workloads*

*1) Parallelization of the Attention Mechanism:* The attention mechanism in transformers, which calculates the importance of each token in a sequence relative to others, is computationally intensive and relies heavily on matrix multiplications. By applying the parallelized block-wise GEMM execution strategy, this CGRA architecture enables significant acceleration of the attention mechanism.

The primary features enabling this parallelized approach include:

- $4 \times 4$ *PE Array Utilization*: The $4 \times 4$ PE array is employed to execute matrix multiplications in parallel, effectively distributing the attention mechanism's computation across multiple PEs. This parallelism reduces the time required to compute attention scores and intermediate transformations.
- *Efficient Data Access through Switchless Interconnect*: Intermediate results generated by the attention mechanism are accessible without delay through the switchless interconnect. This streamlined communication further enhances the speed of GEMM operations by reducing the latency associated with data transfer between PEs. The switchless mesh-torus interconnect simplifies data flow, ensuring that each PE has timely access to the necessary data for ongoing matrix operations.

By optimizing these matrix operations, the architecture achieves a more efficient and responsive execution of the attention mechanism, allowing it to better meet the real-time demands of transformer-based inference tasks.

*2) Application to Edge Scenarios:* The CGRA's design aims to balance high performance with ultra-low-power consumption ($> 1mW$), which is essential for edge applications constrained by energy and processing resources. Transformer models, though powerful, are typically power-intensive, making them challenging to deploy on edge devices.

Key optimizations for edge applications include:

- *Power Efficiency through the Switchless Interconnect*: Traditional interconnects add overhead in both power and latency due to complex switching mechanisms. The switchless mesh-torus interconnect in this CGRA significantly reduces power consumption by eliminating the need for a traditional switching network, which reduces both energy and delay associated with routing data across PEs.
- *Efficient Load/Store Operations*: The $4 \times 2$ MOB optimizes the transfer of data between external memory and on-chip PEs. This efficient LOAD/STORE management minimizes the power required for memory access, reducing the total energy consumed per operation. Such energy optimizations are particularly relevant for battery-powered devices that need to run ML workloads without frequent recharging.

By integrating these architectural enhancements, the CGRA effectively supports transformer-based workloads on ultra-low-power edge devices, providing a viable solution for deploying advanced ML capabilities in constrained environments. This design enables edge devices to execute complex models like transformers more efficiently, paving the way for a new generation of intelligent, real-time applications in distributed, resource-limited settings.

## V. Conclusion

In this paper, the author presented a CGRA architecture to optimize GEMM operations for transformer models deployed at the edge. The design features a heterogeneous array configuration, combining a $4 \times 4$ PE array with ALU and a specialized $4 \times 2$ MOB array for efficient LOAD/STORE handling. This division allows for high parallelism in GEMM computation while reducing data movement and memory access latency, which is crucial for the significant matrix operations characteristic of transformer workloads. The switchless mesh torus interconnect enables low-power, direct communication across PEs, reducing energy consumption and data transfer delays.

The proposed CGRA architecture represents a promising approach for edge deployments, balancing computational performance with energy efficiency to meet the stringent requirements of real-time, on-device transformer inference. By leveraging the CGRA's reconfigurable structure, it offers adaptability to various machine learning tasks beyond transformers, making it a versatile solution for future edge AI applications. This work provides a foundation for further exploration into ultra-low-power CGRA designs, supporting the evolution of intelligent, autonomous edge devices capable of advanced AI processing.

## Acknowledgments

This research was partially funded by the dAIedge project (HORIZON-CL4-2022-HUMAN-02-02, Grant Agreement Number: 101120726).